\documentclass[runningheads]{llncs}

\usepackage[T1]{fontenc}
\usepackage{graphicx}
\usepackage{amsmath}
\usepackage{amssymb}
\usepackage{booktabs}
\usepackage{bbding}

\usepackage{multirow}
\usepackage{url}
\usepackage[hidelinks]{hyperref}

\begin{document}

\title{SphereVAE: Hyperspherical Latent Autoencoders for Robust Autoregressive Speech Representation Modeling}
\titlerunning{SphereVAE for Autoregressive Speech Representation Modeling}

% \author{Anonymous Authors}
% \authorrunning{Anonymous Authors}
% \institute{Anonymous Affiliation}

\author{
Haoyu Zhang\inst{1,2} \and
Jingbin Hu\inst{1} \and
Hanke Xie\inst{1} \and
Qirui Zhan\inst{1} \and
Wenhao Li\inst{1} \and
Ziyu Zhang\inst{1} \and
Xiaming Ren\inst{1} \and
Yue Li\inst{2} \and
Xunyu Zhu\inst{2} \and
Zhipeng Chen\inst{2} \and
Lei Xie\inst{1}\Envelope
}

\authorrunning{H. Zhang et al.}

\institute{
Audio, Speech and Language Processing Group (ASLP@NPU),\\
School of Computer Science,
Northwestern Polytechnical University,
Xi'an, China
\and
Fuxi AI Lab, NetEase Inc.,
Hangzhou, China \\
\email{haoyuz@mail.nwpu.edu.cn}
\email{lxie@nwpu.edu.cn}
}

\maketitle              

\begingroup
\renewcommand{\thefootnote}{\Envelope}
\footnotetext{Indicates the corresponding author.}
\endgroup
\setcounter{footnote}{0}
\begin{abstract}

With the rapid development of speech generation technology, discrete codec representations have been widely used because they provide a stable prediction paradigm. In expressive speech generation, however, the quantization bottleneck of discrete codecs results in information gaps in fine-grained prosody, timbre, pronunciation, and frame-to-frame continuity. Continuous representations (e.g., VAE latents), by eliminating this constraint, have emerged as a more effective alternative for autoregressive modeling. Yet when continuous representations are used as autoregressive prediction targets, prediction errors can accumulate along the generation chain, causing latent drift and degrading long-form stability. To mitigate this problem, we propose SphereVAE, which constrains the VAE latent space to the unit hypersphere. SphereVAE defines a Power Spherical posterior on the hypersphere and regularizes the latent distribution toward a uniform prior, so that information is encoded mainly by directional variation, providing a bounded geometric target for autoregressive prediction and reducing the risk of norm drift. SphereVAE underperforms the standard VAE on reconstruction metrics due to reduced latent freedom. However, when integrated into VoxCPM for zero-shot TTS and long-text generation, it yields lower content error rates with comparable speaker similarity, and shows more stable long-range speaker consistency. These results indicate that an appropriate latent geometric constraint can effectively mitigate autoregressive error accumulation and drift in speech generation.

\end{abstract}

\section{Introduction}

% Recent speech generation systems increasingly rely on learned intermediate representations rather than directly modeling waveforms. 
With the development of large language models, representation-space modeling has become an effective paradigm for text-to-speech synthesis. Compared with directly modeling raw waveforms, learned intermediate representations substantially reduce sequence length and computational cost, while making generation models easier to train. More importantly, in LLM-based speech generation, the quality of these representations largely determines the upper bound of the generated speech, since they carry the acoustic, linguistic, and speaker-related information available to the model. Neural audio codecs and speech tokenizers therefore play a central role by compressing speech into shorter and more regular representation sequences, allowing speech generators to operate in this representation space before reconstructing speech with a decoder. From VALL-E \cite{wang2023vall_e} to the CosyVoice series \cite{du2024cosyvoice,du2024cosyvoice2,du2025cosyvoice3}, the MOSS series \cite{mossttsfamily}, and the Qwen series \cite{hu2026qwen3tts,xu2025qwen3,team2026qwen3}, recent systems commonly use speech tokens or audio tokens as key intermediate units for text-to-speech synthesis, streaming generation, controllable voice cloning, and multimodal spoken interaction. This trend suggests that speech representations for LLM-based generation should not be optimized solely for reconstruction fidelity. Instead, they need to balance acoustic recoverability with the requirements of downstream sequence modeling, including predictability, robustness, and the preservation of task-relevant speech information. This motivates a closer examination of how different speech representation forms trade off modeling stability, reconstruction quality, and expressive capacity.

Discrete codec tokens offer an appealing representation because they provide a bounded prediction space. Each generation step is constrained to a codebook entry or a small set of discrete units, enabling speech generation to be formulated as a classification-based sequence modeling problem. This design improves modeling stability, but it also maps a continuous acoustic space into a finite codebook. Fine-grained prosody, timbre variation, pronunciation details, and frame-to-frame smoothness must all be represented using a finite set of discrete units. Discretization therefore provides a stable modeling interface but may discard fine-grained acoustic information.

Continuous latent representations offer a complementary direction. Unlike discrete tokens, continuous latents do not force each speech frame into a fixed codebook, thereby preserving a more flexible acoustic space for subtle variation. Recent work has explored this direction from different perspectives. DiTAR \cite{jia2025ditar} extends autoregressive speech generation to continuous-representation patches via diffusion-transformer prediction. KALL-E \cite{xia2026kall} directly predicts continuous speech distributions with an autoregressive language model, without relying on a diffusion head. VoxCPM \cite{zhou2025voxcpm,zhou2026voxcpm2} and Dots.tts \cite{lian2026dotstts} further explore continuous speech spaces for autoregressive TTS, demonstrating the potential of prediction-friendly representations beyond discrete tokenization. These studies suggest that continuous speech representations are not meant to replace discrete tokens in all cases, but rather to provide a richer, more fine-grained representation space with greater capacity for expressive generation.

When continuous latents are used in autoregressive generation, the model predicts future representations step by step from historical latents. Unlike discrete token prediction, an error in continuous latent prediction is not merely a wrong class decision. It can appear as a shift in the vector space. A small one-step shift may be tolerable in reconstruction, but in autoregressive generation the previous prediction becomes part of the next-step condition. Local errors may therefore propagate over time, gradually pushing the generated latent sequence away from the training distribution. This issue is particularly critical for speech generation, where the output is a long, temporally ordered sequence and each local acoustic deviation may affect subsequent predictions. Compared with many vision generation settings, speech generation is therefore more sensitive to autoregressive error accumulation, especially in long-form synthesis. 
In this paper, we refer to this problem as latent drift and error accumulation in autoregressive modeling of continuous speech representations.

This problem also shows why continuous speech representations should not be evaluated only by reconstruction metrics. A latent space that reconstructs input speech well may still be difficult to predict over a long horizon. For autoregressive generation, the key question is whether the latent space has a stable geometry that prevents unbounded drift during autoregressive prediction. Conventional VAEs usually model latents in a Euclidean Gaussian space, where both vector norm and direction can carry information. When an autoregressive model predicts such latent variables step by step, errors in magnitude and direction may accumulate and degrade the stability of long-form generation.

To address this problem, we propose SphereVAE: a hyperspherical latent autoencoder that constrains continuous speech representations to a unit sphere and makes autoregressive prediction operate over bounded directional latents. Inspired by hyperspherical latent modeling in visual representation learning and built on the VAE framework for continuous representation learning, SphereVAE replaces the Euclidean Gaussian latent space in a speech representation autoencoder with a unit hyperspherical latent space. This design fixes the latent norm, organizes information mainly through directional variation, and provides a bounded target for autoregressive prediction. Unlike prior uses of spherical latent spaces that primarily study representation geometry or visual generation, our focus is on whether this geometric constraint improves the autoregressive predictability of continuous speech representations. Extensive experiments demonstrate that SphereVAE achieves the lowest content error rates among the VAE variants compared in the VoxCPM-style zero-shot TTS setting. In additional long-text generation experiments, SphereVAE also preserves speaker similarity better than VAE, SemanticVAE, and $\sigma$-VAE, achieving the best overall speaker-similarity retention among the VAE variants compared. Our model and code will be open-sourced.\footnote{\url{https://github.com/ASLP-lab/SphereVAE}} A demo page is available online.
\footnote{\url{https://haoyuzhang3.github.io/SphereVAE_Demo/}}

\section{Related Work}

\subsection{Continuous Speech Representation Learning}

Recent LLM-based speech generation systems increasingly rely on learned intermediate representations rather than directly modeling waveforms or acoustic features. Neural audio codecs, such as VQ-VAE \cite{vq-vae}, DAC \cite{DAC}, and EnCodec \cite{defossez2022encodec}, demonstrate that speech can be compressed into compact representations while preserving information for reconstruction and generation. These models suggest that a tokenizer is not only a compression module, but also an interface that determines what information is preserved for downstream generators.

Most existing speech tokenizers discretize acoustic signals into codebook indices or token sequences. This design provides compact and bounded representations for zero-shot and controllable TTS systems. However, discretization imposes a finite vocabulary on a continuous acoustic space, making fine-grained prosody, timbre, pronunciation, and temporal variations difficult to preserve. Continuous tokenizers avoid hard quantization and provide a richer representation space with greater acoustic capacity.

VAEs provide a natural framework for continuous speech tokenization. A standard VAE \cite{kingma2014autoencoding} learns continuous latent representations through encoder-decoder reconstruction, enabling compact tokens with richer acoustic details than discrete assignments. SemanticVAE \cite{niu2025semanticvae} improves this framework through semantic alignment, while $\sigma$-VAE \cite{sigmaVAE} improves latent sampling by enforcing a fixed latent variance, using an example-level scalar variance shared across channels to prevent variance collapse. These methods improve continuous VAE tokenizers from semantic supervision and latent modeling perspectives, but they do not explicitly address error accumulation during autoregressive prediction of continuous latents.

SphereVAE follows the same continuous-tokenizer direction, but changes the geometry of the latent manifold. We are inspired by Hyperspherical VAE \cite{davidson2018hyperspherical,ke2025spherear} and Power Spherical Distribution \cite{decao2020power}, which show that a unit hypersphere can serve as a principled latent space for VAEs and that probabilistic modeling on the hypersphere can be implemented efficiently. SphereVAE applies this spherical VAE idea to continuous speech representation learning. By fixing the latent norm and organizing information through direction, it aims to preserve the information capacity of continuous representations while giving the latent space a clearer geometric boundary.

\subsection{Autoregressive Modeling of Continuous Representations}

Continuous autoregressive TTS provides another route toward high-fidelity synthesis. Recent work increasingly combines autoregressive planning with diffusion rendering, integrating semantic-prosodic planning and local acoustic rendering into a single generation framework. DiTAR introduces a patch-based generation strategy that combines autoregressive language modeling with diffusion-transformer prediction over speech representation patches, thereby extending continuous autoregressive modeling. VibeVoice \cite{peng2025vibevoice} introduces a hybrid tokenizer that combines acoustic and semantic information into a next-token diffusion framework, offering another way to bridge token-level planning and local acoustic generation.

VoxCPM further explores coarse-to-fine hierarchical modeling for continuous autoregressive TTS, separating higher-level planning from finer acoustic realization while keeping the generation process tokenizer-free. Dots.tts treats the continuous latent space as the modeling target of a TTS foundation model and uses post-training techniques inspired by self-correction to improve generation behavior. Together with related continuous or hybrid AR systems such as GMM-LM \cite{lin2025continuous} and SemaVoice \cite{wang2026semavoice}, these studies show that continuous representations are becoming an active interface for high-quality TTS generation.

Despite this progress, continuous-representation AR TTS still lacks a mature and stable modeling paradigm. The central difficulty is that prediction errors in continuous spaces can accumulate over long horizons in autoregressive generation, degrading audio quality. Existing work mainly improves the generator, alignment strategy, semantic conditioning, or local diffusion prediction module. SphereVAE is orthogonal to these directions: it focuses on the representation prior to its passage to an autoregressive model. The paper asks whether a hyperspherical VAE latent space can reduce drift in continuous speech representations and improve the robustness of long-form generation.

\section{Method}

\subsection{Hyperspherical Latent Space Formulation}
\label{sec:spherevae}

SphereVAE adopts an encoder-decoder architecture for learning speech representations. The encoder maps the input speech to a latent representation, and the decoder reconstructs speech from the latent. Instead of the Euclidean Gaussian posterior used in a standard VAE, SphereVAE constrains the latent to the unit hypersphere $\mathbb{S}^{d-1}$ and models it with a Power Spherical distribution.
\begin{figure}
\centering
\includegraphics[width=0.9\textwidth]{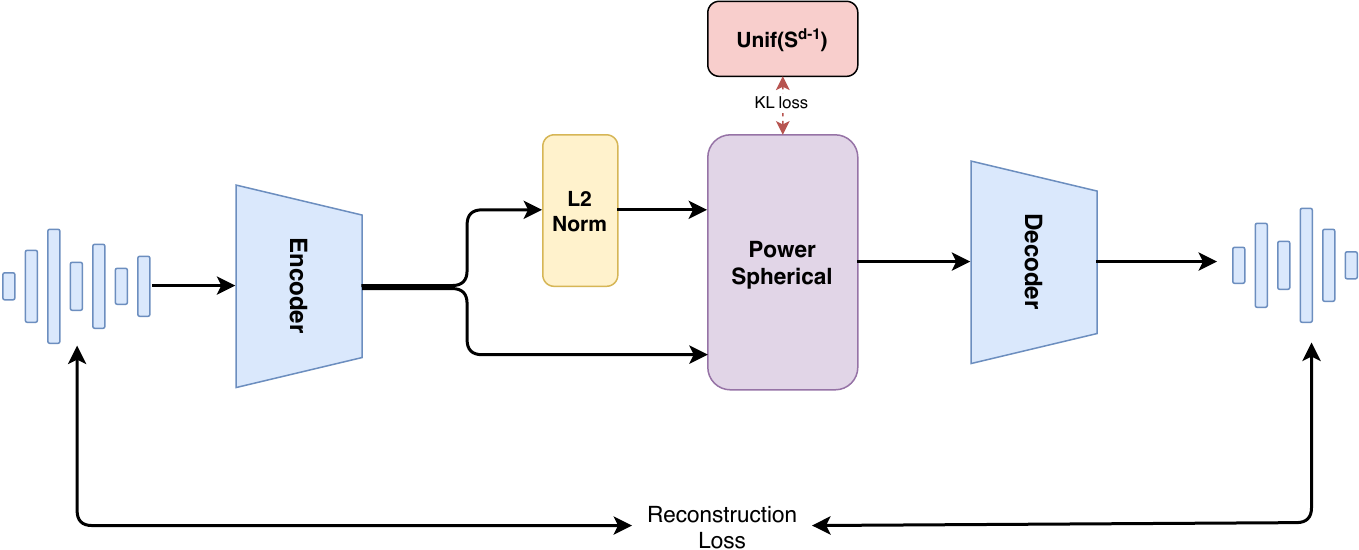}
\caption{Overall architecture of SphereVAE. The encoder maps input speech into latent parameters; the direction branch is L2-normalized to lie on the unit hypersphere, and the Power Spherical posterior is regularized toward the uniform hyperspherical prior. The decoder reconstructs speech from the sampled latent representation.}
\label{fig:svae_architecture}
\end{figure}

Figure~\ref{fig:svae_architecture} summarizes the input-to-output flow of SphereVAE. The input speech is first encoded into hidden features, which are split into a direction branch and a concentration branch. The direction branch is L2-normalized to form the mean direction \(\mu\), while the concentration branch estimates \(\kappa\). These parameters define a Power Spherical posterior over the hypersphere. A sampled latent is then passed to the decoder for waveform reconstruction. During training, the reconstruction objective compares the decoded speech with the input speech, while the KL term regularizes the posterior toward the uniform prior on \(\mathbb{S}^{d-1}\). This architecture differs from a standard Euclidean VAE primarily at the latent bottleneck: the representational channel is free of norm variation, and speech information is organized via bounded directional variation.

Formally, the Power Spherical density on $\mathbb{S}^{d-1}$ takes the form
\begin{equation}
p(z \mid \mu, \kappa) \propto (1 + \mu^\top z)^\kappa, \quad z \in \mathbb{S}^{d-1},
\end{equation}
where \(\mu \in \mathbb{S}^{d-1}\) is the mean direction and \(\kappa > 0\) is the concentration parameter controlling how tightly samples cluster around \(\mu\). The encoder produces a raw direction vector \(h_\mu(x)\) and a scalar \(h_\kappa(x)\) from the encoded features. The direction is L2-normalized as
\begin{equation}
\mu = \frac{h_\mu(x)}{\|h_\mu(x)\|_2},
\end{equation}
and the concentration is obtained via
\begin{equation}
\kappa = \mathrm{softplus}(h_\kappa(x)) + 1.
\end{equation}
This parametrization constrains the concentration to \(\kappa \ge 1\), reducing the risk of an under-concentrated posterior during early training. After sampling a latent vector from the Power Spherical posterior via reparameterization, we rescale it by \(\sqrt{d}\) (where \(d\) is the latent dimension) and pass it to the decoder. The latent regularization is computed as the Kullback--Leibler (KL) divergence from the posterior to the uniform distribution on \(\mathbb{S}^{d-1}\):

\begin{equation}
\mathcal{L}_{\mathrm{KL}} = D_{\mathrm{KL}}\bigl(\mathrm{PowerSpherical}(\mu, \kappa) \,\|\, \mathrm{Uniform}(\mathbb{S}^{d-1})\bigr).
\end{equation}

This design changes how information is represented in the latent space. A standard Gaussian latent can encode differences through both norm and direction. This flexibility aids reconstruction but also leaves an unbounded space for autoregressive drift. SphereVAE enforces a norm, thereby forcing the model to organize speech representations primarily through directional variation. For reconstruction, this constraint may reduce representational freedom. For autoregressive modeling, it provides a clearer geometric boundary and makes it harder for prediction errors to grow through the norm dimension. The trade-off is central to this paper: a spherical latent may not always improve one-step reconstruction, but it may make continuous representation sequences easier to predict.

\subsection{Training Objective}
\label{sec:training}

The training objective combines waveform-level reconstruction, hyperspherical latent regularization, and adversarial supervision. Given an input waveform \(x\), the encoder samples a latent vector from the Power Spherical posterior and the decoder reconstructs \(\hat{x}\). Training alternates between a discriminator step and a generator step. In the discriminator step, the multi-scale STFT discriminator receives the real waveform \(x\) and the detached reconstruction \(\hat{x}\), and learns to distinguish real speech from reconstructed speech. In the generator step, the encoder-decoder is optimized with
\begin{equation}
\mathcal{L}_G = \lambda_{\mathrm{adv}} \,\mathcal{L}_{\mathrm{adv}}^G + \lambda_{\mathrm{fm}} \,\mathcal{L}_{\mathrm{fm}} + \lambda_{\mathrm{recon}} \,\mathcal{L}_{\mathrm{recon}} + \lambda_{\mathrm{KL}} \,\mathcal{L}_{\mathrm{KL}},
\end{equation}
where the four terms are defined as follows.

\textbf{Reconstruction loss} \(\mathcal{L}_{\mathrm{recon}}\). Following DAC, we use a multi-scale mel-spectrogram loss over multiple window lengths. It sums the \(L_1\) distance between the mel-spectrograms of the ground-truth waveform \(x\) and the reconstruction \(\hat{x}\):
\begin{equation}
\mathcal{L}_{\mathrm{recon}} = \sum_{w \in \mathcal{W}} \left\| \mathrm{Mel}_w(x) - \mathrm{Mel}_w(\hat{x}) \right\|_1 .
\end{equation}
This term captures both the global spectral envelope and local time-frequency detail and serves as the primary reconstruction signal.

\textbf{KL regularization} \(\mathcal{L}_{\mathrm{KL}}\). This is the KL divergence from the Power Spherical posterior to the uniform prior on \(\mathbb{S}^{d-1}\), as defined in Section~\ref{sec:spherevae}. It keeps the posterior tied to the intended hyperspherical geometry while still allowing the encoder to use directional concentration to represent speech-dependent variation.

\textbf{Adversarial loss} \(\mathcal{L}_{\mathrm{adv}}^G\). Separately from the mel-spectrogram reconstruction loss, a multi-scale STFT discriminator following the EnCodec design processes \(x\) and \(\hat{x}\) at several STFT resolutions. The generator adversarial loss encourages reconstructed speech to be classified as real, while the discriminator is trained with its own objective
\begin{equation}
\mathcal{L}_{\mathrm{adv}}^D = \mathcal{L}_{\mathrm{disc}}(D(x), D(\hat{x})).
\end{equation}
This term complements the reconstruction loss by penalizing perceptual artifacts that may not be fully captured by pointwise spectral distances.

\textbf{Feature matching loss} \(\mathcal{L}_{\mathrm{fm}}\). The L1 distance is computed between the intermediate discriminator feature maps for \(x\) and \(\hat{x}\):
\begin{equation}
\mathcal{L}_{\mathrm{fm}} = \sum_{\ell} \left\| D_{\ell}(x) - D_{\ell}(\hat{x}) \right\|_1 .
\end{equation}
Here, \(D_{\ell}(\cdot)\) denotes an intermediate feature map of the discriminator. This term stabilizes GAN training and encourages the decoder to reproduce the multi-resolution acoustic characteristics captured by the discriminator.

\section{Experiments}

\subsection{Experimental Setup}

\subsubsection{Model Setup}

SphereVAE follows the Mimi \cite{defossez2024moshi} architecture, uses SEANet as the encoder--decoder backbone, and replaces the unconstrained Euclidean latent space with a unit hyperspherical latent space. The input waveform is sampled at 24~kHz and processed by convolutional downsampling layers with rates [8, 6, 5, 4], yielding a total downsampling factor of 960 and a 25~Hz latent sequence. The latent dimension is 64. The decoder mirrors the encoder, using symmetric upsampling rates, and reconstructs waveform-level speech from the continuous latent sequence. The discriminator follows the EnCodec discriminator design. The model is trained with the GAN-VAE hybrid objective defined in Section~\ref{sec:training}, using loss weights $\lambda_{\mathrm{adv}} = 1.0$, $\lambda_{\mathrm{msspec}} = 15$, $\lambda_{\mathrm{fm}} = 1.5$, and $\lambda_{\mathrm{KL}} = 0.01$. Both the generator and the discriminator are optimized with AdamW using $\beta = (0.5, 0.9)$, a learning rate of $2 \times 10^{-4}$, and a warmup-cosine schedule. SphereVAE is trained for 500k steps on 12-second audio segments with a batch size of 32.

VAE, SemanticVAE, and $\sigma$-VAE are trained as controlled baselines using the same model architecture and training hyperparameters. VAE denotes the standard continuous VAE baseline. SemanticVAE modifies the representation through semantic regularization, while $\sigma$-VAE improves latent sampling by enforcing a fixed latent variance. In this setup, the comparison isolates the effect of the hyperspherical latent geometry from other training factors.

\subsubsection{Zero-shot TTS Model Setup}

To evaluate the downstream generation performance of different VAE representations, we use VoxCPM as the autoregressive backbone in a zero-shot TTS setting.  The autoregressive backbone predicts latent sequences rather than waveform samples directly, and the generated latents are converted back to speech by the corresponding acoustic decoding path. The VoxCPM model is trained at a 24~kHz sample rate on four GPUs, with a per-device batch size of 16 and a single gradient accumulation step. Training runs for 500k iterations with a learning rate of 1e-4, weight decay of 0.01, 2000 warmup steps, and a maximum batch-token budget of 8192. The diffusion loss and stop-token loss are both weighted by 1.0.

\subsubsection{Training Datasets}

The VAE representation models and the VoxCPM models are both trained on Emilia ZH and EN subsets \cite{he2024emilia}.

\subsubsection{Evaluation}

Tokenizer reconstruction performance is evaluated on LibriSpeech-PC test-clean \cite{meister2023librispeech} using STOI \cite{taal2010stoi}, PESQ-WB \cite{rix2001pesq}, MCD \cite{kubichek1993mcd}, UTMOS \cite{saeki2022utmos}, SIM, and WER. SIM measures speaker similarity between the reconstructed and original audio using WeSpeaker \cite{wang2023wespeaker}, while WER measures the word error rate of reconstructed speech using Whisper \cite{radford2023robust}. Zero-shot TTS performance is evaluated on SeedTTS-eval \cite{anastassiou2024seed} with WER and SIM. Long-text robustness is measured on Long-TTS-Eval from MGM-OMNI \cite{wang2025mgmomni}: overlong texts are shortened below 200 Chinese characters when needed, prompt speech is sampled from SeedTTS-eval, and generated speech is segmented every 3 seconds for prompt-to-segment speaker-similarity comparison. We further use t-SNE \cite{vandermaaten2008tsne} and PCA \cite{jolliffe2016pca} to analyze how the hyperspherical constraint changes the organization of the continuous latent space.

\subsection{VAE Reconstruction Results}

Table~\ref{tab:vae_reconstruction} reports the reconstruction results on the LibriSpeech-PC test-clean set. VAE leads on the reconstruction-oriented metrics, while SemanticVAE gives the highest SIM score. SphereVAE remains within a usable reconstruction range under the hyperspherical constraint. This result supports the intended positioning of SphereVAE: its main goal is not to maximize one-step reconstruction quality, but to provide a constrained continuous latent space for autoregressive generation.

\begin{table}
\caption{Reconstruction results on LibriSpeech-PC test-clean. Arrows indicate whether higher or lower values are better; best scores are bolded.}
\label{tab:vae_reconstruction}
\centering
\scriptsize
\begin{tabular*}{0.96\textwidth}{@{\extracolsep{\fill}}lcccccc@{}}
\toprule
Model & STOI$\uparrow$ & PESQ-WB$\uparrow$ & MCD$\downarrow$ & UTMOS$\uparrow$ & SIM$\uparrow$ & WER$\downarrow$ \\
\midrule
VAE & \textbf{0.9707} & \textbf{3.470} & \textbf{1.996} & \textbf{3.966} & 0.772 & \textbf{0.038} \\
SphereVAE & 0.9574 & 3.009 & 2.443 & 3.580 & 0.673 & 0.042 \\
SemanticVAE & 0.9689 & 3.411 & 2.043 & 3.913 & \textbf{0.780} & \textbf{0.038} \\
$\sigma$-VAE & 0.9656 & 3.370 & 2.111 & 3.847 & 0.752 & 0.040 \\
\bottomrule
\end{tabular*}
\end{table}

\vspace{-0.7em}

The gap between VAE and SphereVAE is consistent with the geometric constraint imposed on the latent space. By restricting latent vectors to the unit hypersphere, SphereVAE reduces part of the reconstruction freedom available to an unconstrained Euclidean VAE. The following generation experiments therefore evaluate whether this constrained geometry improves autoregressive predictability and stability of generation.

\subsection{Zero-shot TTS Model Evaluation}

The Zero-shot TTS Model evaluation tests whether the continuous latents are easier to predict in an autoregressive zero-shot TTS setting. SphereVAE achieves a WER of 5.305\% on the EN test set and a CER of 1.141\% on the ZH test set, yielding the lowest content error rates among the four VAE representations compared. VAE is the closest baseline on EN with 5.591\% WER, while SemanticVAE is the closest baseline on ZH with 1.241\% CER. These results indicate that the spherical latent is more favorable for content prediction in the current zero-shot TTS setting.

\begin{table}
\caption{Zero-shot TTS Model results on SeedTTS-eval. Lower WER is better and higher SIM is better.}
\label{tab:voxcpm_seed}
\centering
\begin{tabular*}{0.86\textwidth}{@{\extracolsep{\fill}}lcccc@{}}
\toprule
\multirow{2}{*}{model} & \multicolumn{2}{c}{EN} & \multicolumn{2}{c}{ZH} \\
\cmidrule(lr){2-3} \cmidrule(lr){4-5}
 & WER$\downarrow$ & SIM$\uparrow$ & CER$\downarrow$ & SIM$\uparrow$ \\
\midrule
SphereVAE & \textbf{5.305} & 0.654 & \textbf{1.141} & \textbf{0.733} \\
VAE & 5.591 & \textbf{0.655} & 1.327 & 0.731 \\
SemanticVAE & 7.702 & 0.654 & 1.241 & 0.731 \\
$\sigma$-VAE & 9.442 & 0.534 & 4.621 & 0.675 \\
\bottomrule
\end{tabular*}
\end{table}

The SIM differences among SphereVAE, VAE, and SemanticVAE are small. VAE is slightly higher on the EN test set SIM, while SphereVAE is slightly higher on the ZH test set SIM; $\sigma$-VAE is clearly lower on both language subsets. Taken together with the reconstruction results, SphereVAE does not gain its advantages through stronger reconstruction quality. A more plausible explanation is that the constrained latent geometry reduces the difficulty of autoregressive prediction. The benefit of the hyperspherical constraint therefore lies mainly in autoregressive predictability and generation stability, rather than in directly optimizing all acoustic similarity metrics.

\subsection{Long-Text Inference Evaluation}

The long-text inference experiment directly tests error accumulation in longer generation. The goal is not merely to compare the average SIM across short utterances, but to examine whether speaker drift weakens as generation length increases. If autoregressive prediction over continuous latent variables suffers from error accumulation, later segments should show a persistent or abrupt decrease in speaker similarity. In the common 0--27~s analysis window, the updated segment-wise curve shows that SphereVAE maintains higher similarity in most comparable segments, especially over the late 18--27~s range. This trend suggests that the hyperspherical latent space helps reduce speaker drift and provides more direct evidence for mitigating long-horizon error accumulation in continuous AR speech representation model.

\begin{table}
\caption{Long-text speaker similarity retention on Long-TTS-Eval over the common 0--27~s range.}
\label{tab:long_tts_sim}
\centering
\begin{tabular*}{0.86\textwidth}{@{\extracolsep{\fill}}lrrrrr@{}}
\toprule
Model & F\_SIM$\uparrow$ & L\_SIM$\uparrow$ & M\_SIM$\uparrow$ & $\Delta$SIM$\downarrow$ & SIM$_{late}\uparrow$ \\
\midrule
VAE & 0.653 & 0.410 & 0.567 & 0.243 & 0.494 \\
SphereVAE & \textbf{0.667} & \textbf{0.447} & \textbf{0.591} & 0.220 & \textbf{0.524} \\
SemanticVAE & 0.655 & 0.404 & 0.562 & 0.251 & 0.480 \\
$\sigma$-VAE & 0.571 & 0.392 & 0.533 & \textbf{0.179} & 0.478 \\
\bottomrule
\end{tabular*}
\end{table}

Table~\ref{tab:long_tts_sim} summarizes speaker similarity across the common evaluation range, with each model having at least 10 valid pairs in the first 9 3-second segments (0--27~s). After 27~s, the number of valid generated utterances drops sharply for some models, making the 27--30~s segment statistically unreliable; therefore, the main results are reported only up to segment 8. $\mathrm{F\_SIM}$ and $\mathrm{L\_SIM}$ denote the SIM scores of the first and last segments in this range, $\mathrm{M\_SIM}$ is the average SIM over all nine segments, $\Delta\mathrm{SIM}$ measures the first-to-last decrease, and $\mathrm{SIM}_{late}$ is the average SIM over the last three segments (18--27~s). For each model and segment, we also record the effective number of valid pairs, the segment mean, the sample standard deviation, and the 95\% confidence interval computed from the reported segment variance. SphereVAE achieves the highest $\mathrm{M\_SIM}$ and $\mathrm{SIM}_{late}$, indicating stronger retention over longer generations.

\begin{figure}
\centering
\includegraphics[width=0.8\textwidth]{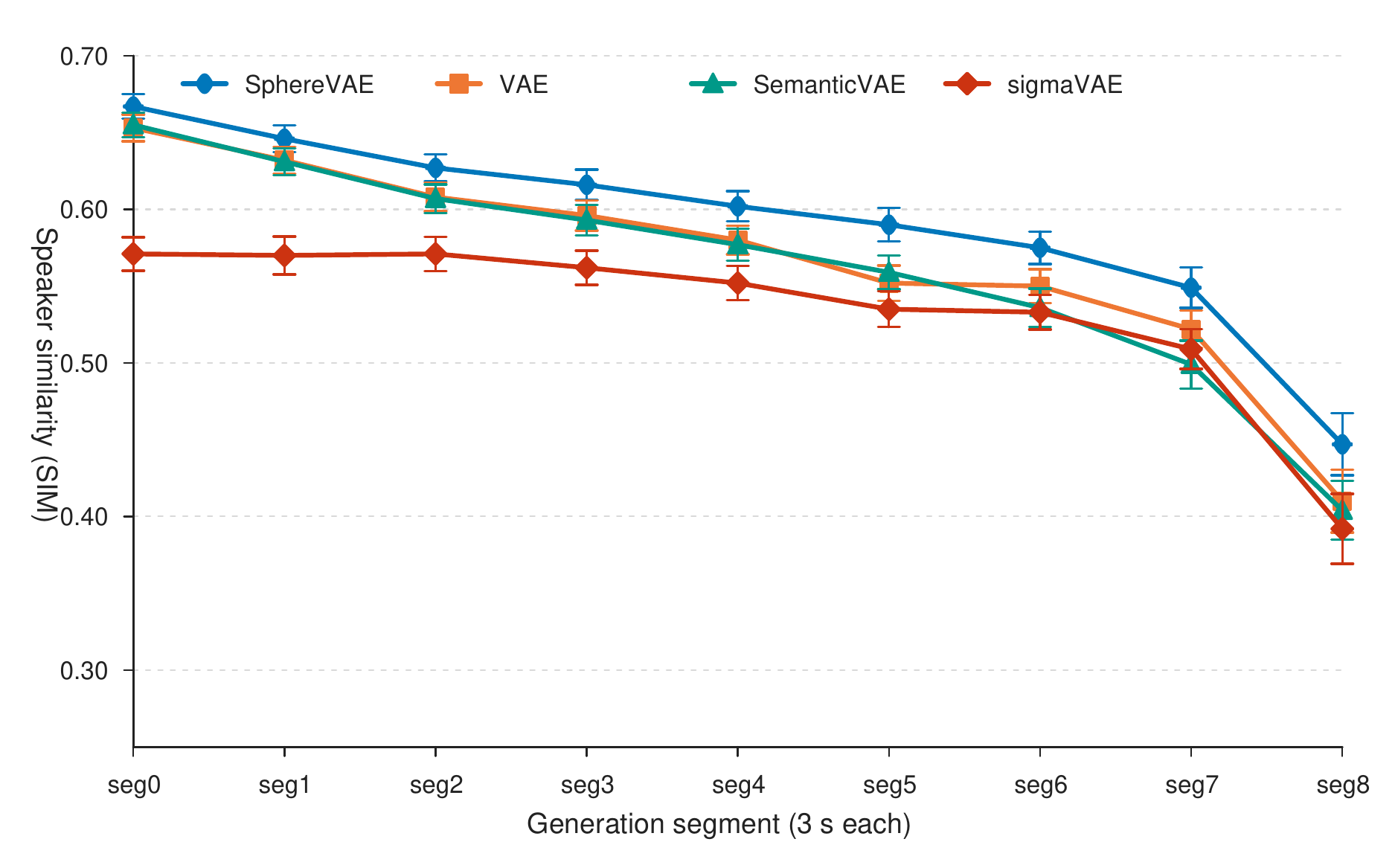}
\caption{Segment-wise speaker similarity curves on Long-TTS-Eval. The generated audio is segmented every 3 seconds, and each segment is compared with the prompt speech.}
\label{fig:long_tts_sim}
\end{figure}

\vspace{-1.3em}

Although $\sigma$-VAE has a smaller first-to-last drop, its first-segment SIM and mean SIM are both lower than the other models. This is better interpreted as a smaller decline from a lower similarity level, rather than stronger long-text stability. Considering mean SIM, late-segment SIM, and first-to-last drop together, SphereVAE better matches the goal of long-text inference: it preserves higher speaker similarity while maintaining a smaller decline than the standard VAE and SemanticVAE.

\subsection{Latent Distribution Analysis with t-SNE and PCA}

\begin{figure}
\centering
\begin{minipage}{0.29\textwidth}
\centering
\includegraphics[width=\textwidth]{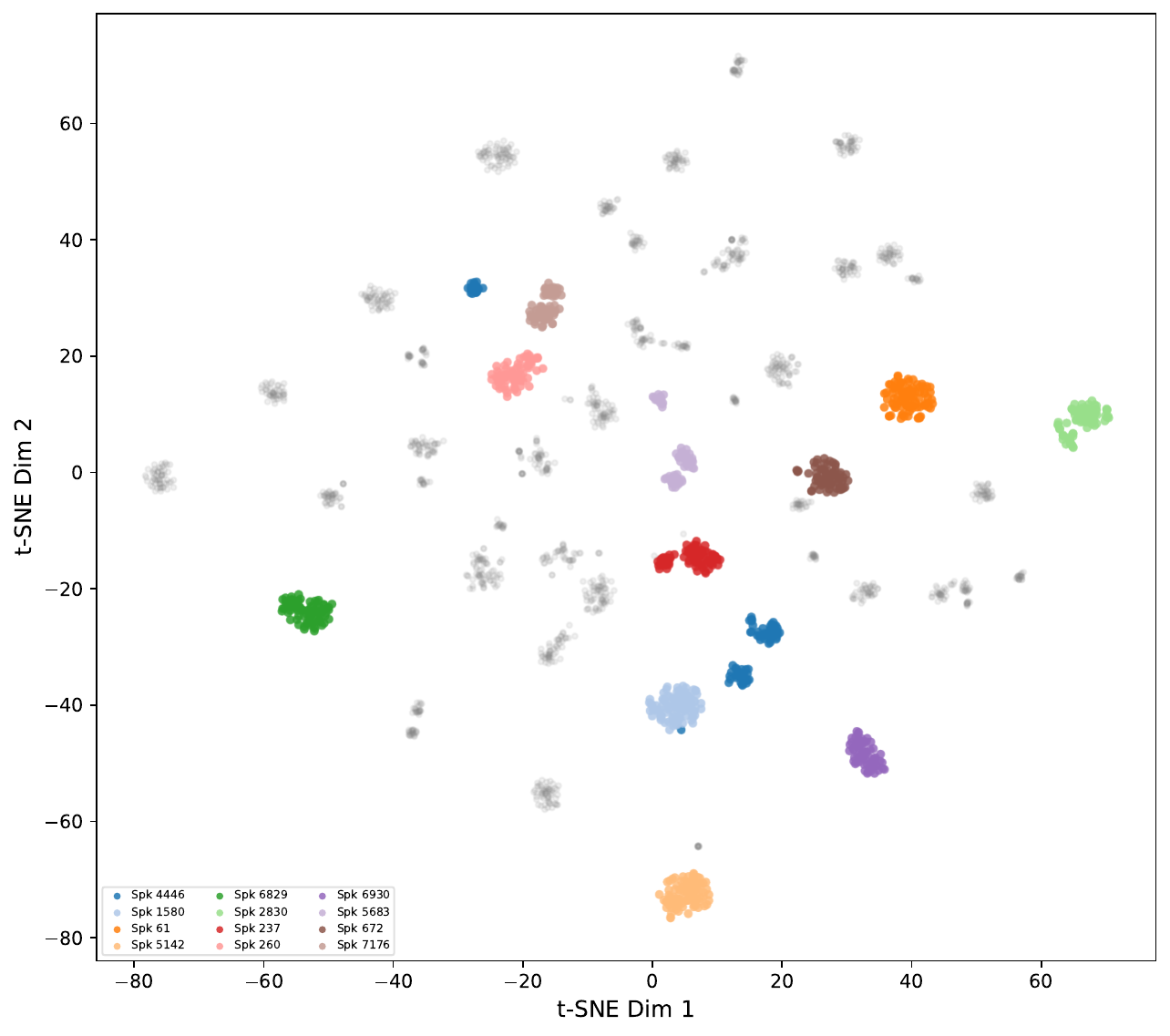}\\
\textbf{(a)} SphereVAE t-SNE
\end{minipage}
\hfill
\begin{minipage}{0.29\textwidth}
\centering
\includegraphics[width=\textwidth]{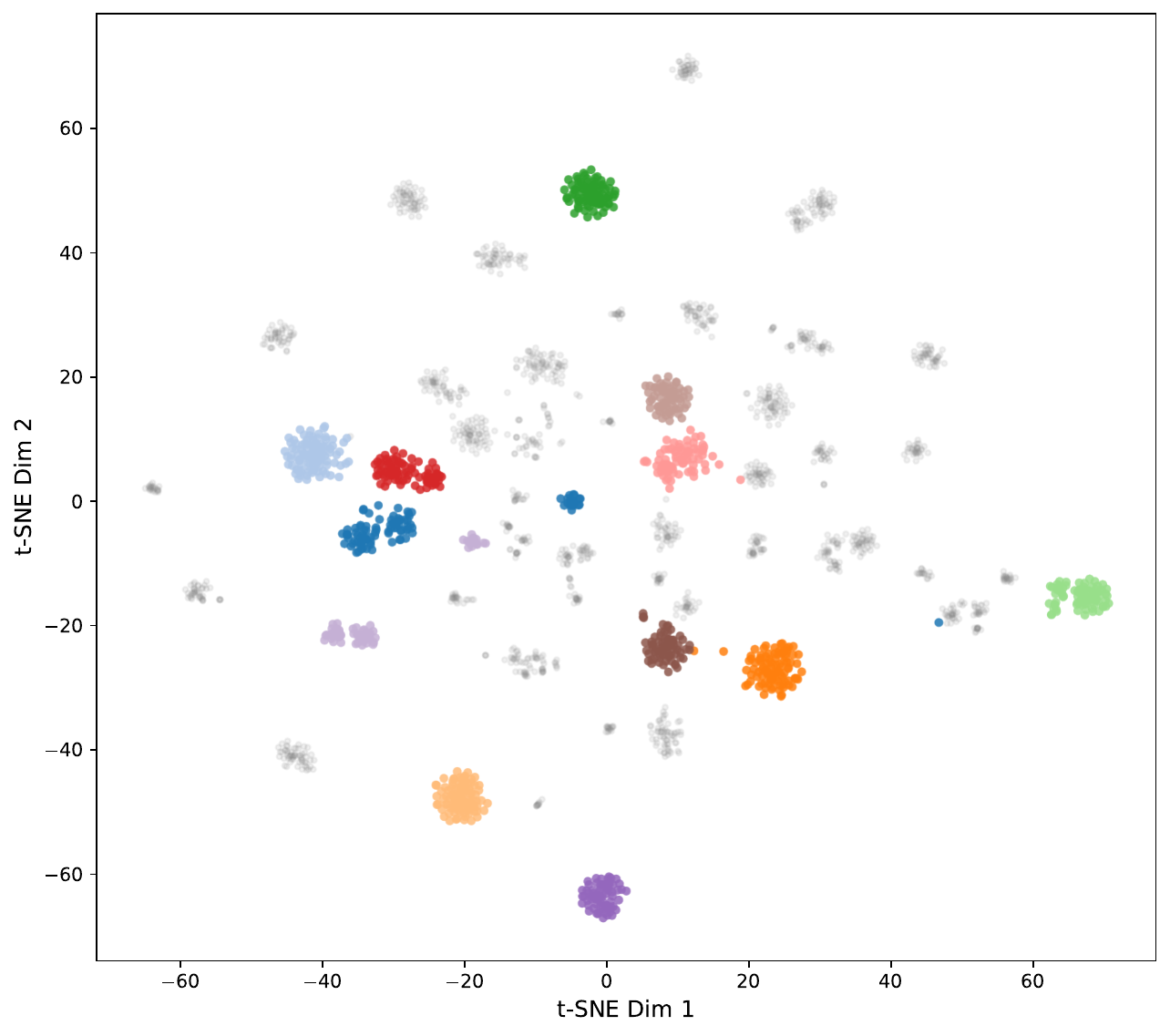}\\
\textbf{(b)} VAE t-SNE
\end{minipage}
\hfill
\begin{minipage}{0.29\textwidth}
\centering
\includegraphics[width=\textwidth]{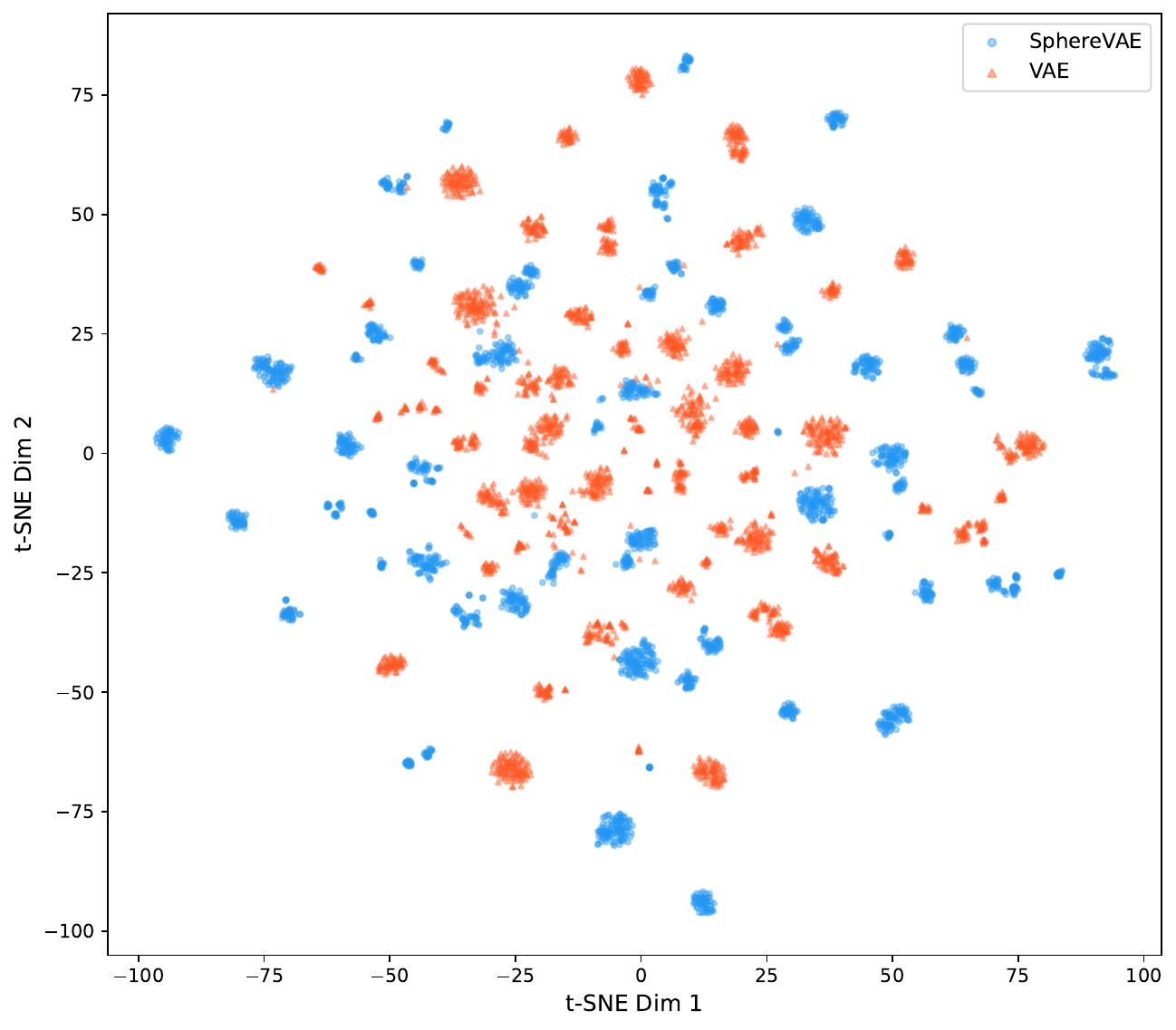}\\
\textbf{(c)} Joint t-SNE
\end{minipage}

\vspace{0.1em}
\makebox[\textwidth][c]{%
\begin{minipage}{0.29\textwidth}
\centering
\includegraphics[width=\textwidth]{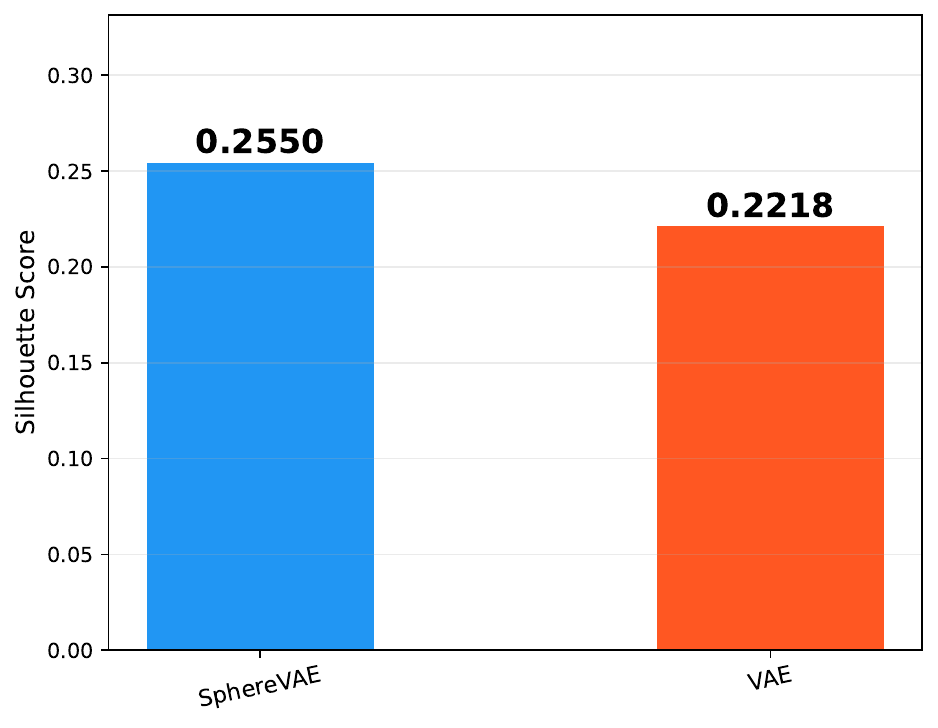}\\
\textbf{(d)} Silhouette
\end{minipage}
\hspace{0.04\textwidth}
\begin{minipage}{0.29\textwidth}
\centering
\includegraphics[width=\textwidth]{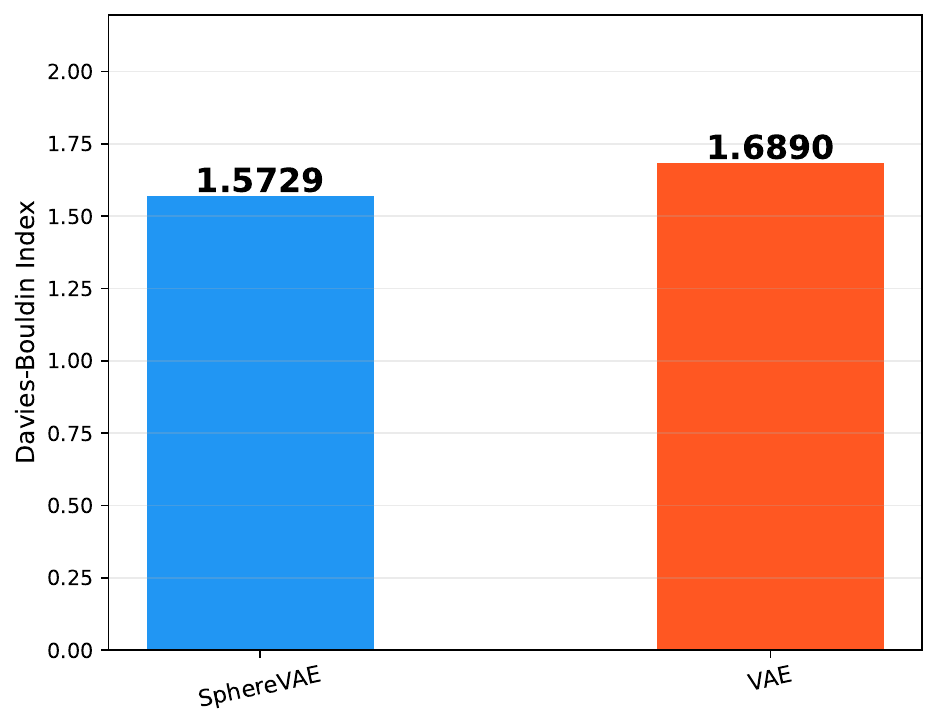}\\
\textbf{(e)} Davies--Bouldin
\end{minipage}
}
\caption{Latent-space comparison between SphereVAE and standard VAE on LibriSpeech-PC test-clean. (a) SphereVAE speaker-colored t-SNE, (b) VAE speaker-colored t-SNE, (c) joint t-SNE overlay, (d) Silhouette Score, and (e) Davies--Bouldin Index.}
\label{fig:tsne_latent}
\end{figure}

To explain the effect of the hyperspherical constraint on the latent distribution, we analyze the latents of SphereVAE and standard VAEs using t-SNE and PCA. For the t-SNE analysis, we use the LibriSpeech-PC test-clean set, comprising 2,620 utterances from 40 speakers. Each utterance-level representation is obtained by mean-pooling the 64-dimensional latent sequence over time before dimensionality reduction. Figure~\ref{fig:tsne_latent} compares the two latent spaces using speaker-colored t-SNE projections, a joint t-SNE overlay, and clustering metrics, including Silhouette Score~\cite{rousseeuw1987silhouettes} and Davies--Bouldin Index~\cite{davies1979cluster}.

The clustering metrics consistently favor SphereVAE. Compared with the standard VAE, SphereVAE shows stronger speaker clustering, lower cluster overlap, and a smaller intra- and inter-distance ratio. These results indicate a more compact within-speaker structure and clearer between-speaker separation in the mean-pooled latent space, suggesting that SphereVAE learns a compact, better-organized latent manifold.

\vspace{-1.3em}

The PCA result provides a complementary view. SphereVAE has a higher cumulative explained variance than VAE under the same number of principal components. This means that SphereVAE concentrates more information in fewer dominant directions, while the standard VAE distributes information more broadly. This observation is consistent with the method design. Since the latent is constrained to the unit hypersphere, the model cannot rely on arbitrary norm variation and tends to organize useful information along major directions. This may reduce reconstruction freedom, but for an autoregressive model, it is easier to predict a bounded directional manifold than an unconstrained sequence of Euclidean vectors.

\begin{figure}
\centering
\includegraphics[width=0.55\textwidth]{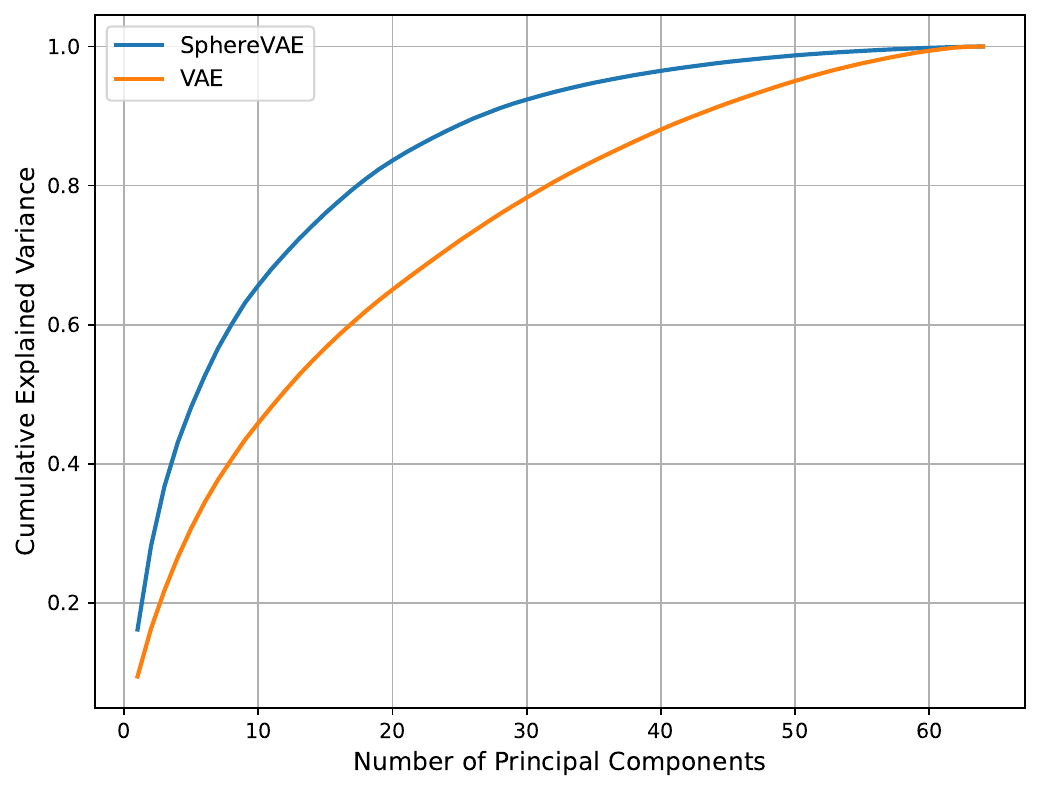}
\caption{PCA cumulative explained variance comparison between SphereVAE and standard VAE latent features.}
\label{fig:pca_variance}
\end{figure}

These qualitative results offer a geometric interpretation consistent with the main hypothesis: the hyperspherical constraint concentrates representational information and yields a structurally distinct latent organization compared with the Euclidean VAE. Quantitatively linking geometric properties of the latent space to autoregressive drift metrics would further strengthen this interpretation.

\section{Conclusion}

In this paper, we investigate the accumulation of error in autoregressive modeling of continuous speech representations and propose SphereVAE, which constrains the VAE latent space to the unit hypersphere. By fixing the latent norm and organizing information through directional variation, SphereVAE provides a bounded geometric target for autoregressive prediction. Experiments show that SphereVAE does not aim to maximize one-step reconstruction quality; instead, it improves downstream generation behavior, achieving the lowest SeedTTS-eval content error rates among VAE, SemanticVAE, and $\sigma$-VAE while keeping speaker similarity comparable to the strongest baselines. It also shows a stronger retention of speaker similarity in long-text inference. PCA and t-SNE analyses further suggest that the hyperspherical constraint changes latent-space organization by concentrating representational variation into more dominant directions. These results indicate that latent-space geometry is a meaningful design axis for continuous speech representations, and future work will develop quantitative drift metrics and evaluate the approach under longer and more diverse generation conditions.

\bibliographystyle{splncs04}

\bibliography{references}

\end{document}